\documentclass[12pt]{article}

\usepackage{amsmath}
\usepackage{amssymb}
\usepackage{amsfonts}
\usepackage{epsfig}
\usepackage{float}
\usepackage[a4paper]{geometry}

\def\d{{\rm d}}
\def\i{{\rm i}}
\def\df#1#2{\frac{{\rm d}#1}{{\rm d}#2}} % ordinary derivative
\def\pd#1#2{\frac{\partial#1}{\partial#2}} % partial derivative
\def\fd#1#2{\frac{\delta#1}{\delta#2}} % functional derivative

\def\H{H}
\def\HH{{\cal H}}
\def\h{h}

\def\ave#1{\left\langle#1\right\rangle}

\def\ang#1#2{\left\langle#1,#2\right\rangle}

\def\sch{Schr\"odinger}
\def\abs#1{\left|#1\right|}

\begin{document}

\title{{Nonlinear parametric quantization of gravity and\\ cosmological models}}
%\title{On nonlinear parametric quantization of gravity and cosmological models}

\author{Charles Wang\\
Department of Physics, Lancaster University\\
Lancaster LA1 4YB, {England}\\
E-mail: c.wang@lancaster.ac.uk}

\date{}

\maketitle

\abstract{A generalization of the recently formulated nonlinear quantization
of a parameterized theory is presented in the context of quantum gravity.
The parametric quantization of a Friedmann universe with a massless scalar field
is then considered in terms of analytic solutions of the resulting evolution equations.}

\vskip 7mm


\vskip 10mm

%\noindent
In a recent paper \cite{wang_2003}
the quantization of a physical system with
finite degrees of freedom subject to a Hamiltonian constraint
was re-examined. A new scheme, called parametric quantization,
was introduced in that work advocating the concept of
treating a chosen time variable as a
constrained classical variable coupled
to the other dynamical variables to be quantized. The approach
was motivated by the
cosmological approach to quantum gravity \cite{misner}
whose dynamical structure is analogous to that of
the parameterized theory of a relativistic particle.
In an $(n+1)$-dimensional pseudo-Riemannian spacetime with a
metric tensor $\gamma_{\mu\nu}(q^\lambda)$
in coordinates $q^\mu,\; (\mu,\nu = 0,1,\cdots,n)$
the dynamics of such a particle subject to a potential $V(q^\lambda)$
is generated
by a Lagrangian of the following generic form:
\begin{equation}\label{L}
L(q^\mu, \dot{q}^\mu, N) = \frac1{2N}\gamma_{\mu\nu} \dot{q}^\mu \dot{q}^\nu - N V.
\end{equation}
Here $q^\mu(t)$ describe the particle trajectory parameterized by $t$, whose
choice is related to the function $N(t)>0$, and
$\dot{q}^\mu := \df{\dot{q}^\mu}{t}$.
Classically the particle's motion is governed by the  Lagrange equations
\begin{equation}\label{}
\df{}t\left(\pd{L}{\dot{q}^\mu}\right) - \pd{L}{{q}^\mu} = 0
\end{equation}
subject to the constraint
\begin{equation}\label{HH}
\HH(q^\mu, \dot{q}^\mu, N) := -\pd{L}{N} =   \frac1{2N^2}\gamma_{\mu\nu} \dot{q}^\mu \dot{q}^\nu + V = 0.
\end{equation}
The equivalent canonical description is obtained from the
``constrained Hamiltonian'':
\begin{equation}\label{H}
\H(q^\mu, p_\mu, N)
{:=}
\dot{q}^\mu p_\mu - L = N \HH
\end{equation}
in terms of the conjugate momenta $p_\mu := \pd{L}{\dot{q}^\mu} = \frac1{N}\gamma_{\mu\nu} \dot{q}^\nu$
where $\HH$ can be re-express as
\begin{equation}\label{}
\HH(q^\mu, p_\mu) = \frac1{2}\gamma^{\mu\nu} p_\mu p_\nu + V.
\end{equation}
The canonical equations of motion are given by
\begin{align}
  \df{{q}^\mu}{t} &= \pd{\H}{p_\mu},\quad
  \df{{p}_\mu}{t} = -\pd{\H}{q^\mu}
\end{align}
subject to the Hamiltonian constraint
\begin{equation}\label{}
\HH(q^\mu, p_\mu)=0.
\end{equation}

A quantization scheme may be set up by choosing one of the
variables, say $q^0$ and its conjugate momentum $p_0$, as classical variables that
interact with
the remaining quantized variables $q^a,\; (a=1,2,\cdots, n)$
in a semi-classical fashion \cite{wang_2003}:
\begin{equation}\label{}
\i \pd{{\psi}}{t} = \hat{\H} \psi
\label{relweq}
\end{equation}
\begin{eqnarray}
\df{q^0}{t} &=& \ave{\pd{\hat{\H}}{p_0}}, \quad  \df{p_0}{t} = - \ave{\pd{\hat{\H}}{q^0}}
\label{reldott}
\end{eqnarray}
\begin{equation}
\ave{\hat{\HH}} = 0
\label{relaveH}
\end{equation}
where $\psi = \psi(q^a,t)$ and $\ave{\hat{O}}$ denotes the expectation value of
any operator $\hat{O}$. \footnote{Units in which $c = \hbar = 16 \pi G = 1$ are adopted throughout.}
The operators $\hat{\H}$ and   $\hat{\HH}$
are obtained by substituting $p_a \rightarrow \hat{p}_a := -\i \pd{}{q^a}$
into ${\H}$ and   ${\HH}$ respectively (followed by a suitable factor ordering.)
Equations
\eqref{relweq}, \eqref{reldott} and \eqref{relaveH}
can be cast into a more compact form that will prove to be advantageous
in generalizing parametric quantization to a field theoretical framework.
This reformulation is done by constructing
an ``unconstrained Hamiltonian'':
\begin{equation}\label{h}
\h(q^a, p_a, q^0, \dot{q}^0, N)
:=
\dot{q}^a p_a - L
=
\H - \dot{q}^0\pd{L}{\dot{q}^0}
\end{equation}
using a partial Legendre transformation of $L$
by leaving out the ``$\dot{q}^0 \rightarrow p_0$'' transform
as performed in \eqref{H}. With this the classical
motion obeys the following canonical-type equations
\begin{align}\label{can0}
  \df{{q}^a}{t} &= \pd{\h}{p_a},\quad
  \df{{p}_a}{t} = -\pd{\h}{q^a}
\end{align}
plus the constraint-like equation
\begin{equation}\label{relh0}
\HH(q^a, p_a, q^0, \dot{q}^0, N)=\frac{1}{N}\left(\h+\dot{q}^0\pd{L}{\dot{q}^0}\right)=0.
\end{equation}

Accordingly, in terms of the operators $\hat{\h}(q^a, \hat{p}_a, q^0, \dot{q}^0, N)$ and
$\hat{\HH}(q^a, \hat{p}_a, q^0, \dot{q}^0, N)$ obtained
by substituting $p_a \rightarrow \hat{p}_a$ into $\h$ and ${\HH}$
given in \eqref{h} and \eqref{relh0},
the system of equations  \eqref{relweq},  \eqref{reldott} and  \eqref{relaveH}
now becomes
\begin{equation}
\i \pd{{\psi}}{t} = \hat{\h} \psi
\label{relweq0}
\end{equation}
\begin{equation}
\ave{\hat\HH} = 0.
\label{relaveH0}
\end{equation}
For any given positive $N(t)$,
equations \eqref{relweq0} and \eqref{relaveH0}
constitute a system of evolution for the wavefunction
$\psi(q^a,t)$ and classical variable $q^0(t)$
whose initial data may be specified arbitrarily. It is evident that the
presence of \eqref{relaveH0} makes the quantum theory {\em nonlinear}.
(See, for instance, \cite{kibble_1978, Weinberg 1989a} for other examples of nonlinear quantum theories.)

This formulation suggests a
parametric quantization of gravity
whose formal treatment can be outlined as follows.
Start from the standard ADM Lagrangian for general relativity \cite{Carlip_2001}:
\begin{equation}\label{}
  L[g_{ij}, \dot{g}_{ij}, N_\mu]
  {:=}
  \int  N \sqrt{g}\left(K_{i j} K^{i j} - K^2 +  R \right) \d^3 x
\end{equation}
in terms of the 3-metric $g_{ij}(x)=g_{ij}(x^k,t)$, $g = \det(g_{ij})$,
lapse function $N(x) = N_0(x)$, shift vector $N_i(x)$,
extrinsic curvature $K_{i j}$, $K = g^{ij}K_{i j}$ and intrinsic scalar curvature $R$
of the evolving 3-geometry. ($\mu=0,1,2,3; i,j,k=1,2,3$.)
A natural way of isolating the ``true'' gravitational degrees of freedom
is to transform from $g_{ij}(x)$ to a set of
embedding variables: $\vartheta^\mu(x), \mu = 0,1,2,3$,
and unconstrained variables: $\varphi^{{r}}(x), {{r}}=1,2$ \cite{kuchar_1992}.
Here
$\vartheta^0$ specifies time slicing and $\vartheta^a$ sets spatial coordinate condition.
We then have $L = L[\varphi^{{r}}, \varpi_{{r}}, \vartheta^\mu, \dot{\vartheta}^\mu, N_\mu]$.
The idea now is to regard $\vartheta^\mu$ as constrained classical variables
coupled ``semi-classically'' to the quantized true degrees of freedom carried by $\varphi^{{r}}$.

In view of the discussions above we introduce the
unconstrained Hamiltonian
\begin{align}\label{}
  h[\varphi^{{r}}, \varpi_{{r}}, \vartheta^\mu, \dot{\vartheta}^\mu, N]
  &:= \int \dot{\varphi}^{{r}} \varpi_{{r}}  \d^3 x - L
\end{align}
with the conjugate momenta
\begin{equation}\label{pr}
  \varpi_{{r}}(x) := \fd{L}{\dot{\varphi}^{{r}}(x)}.
\end{equation}
{
Further, we may derive the
`super-Hamiltonian' $\HH = \HH^0$ and `super-momenta' $\HH^i$  as follows:
\begin{equation}\label{}
\HH^\mu[\varphi^{{r}}, \varpi_{{r}}, \vartheta^\nu, \dot{\vartheta}^\nu, N_\nu; x)
:= \fd{h}{N_\mu(x)}
\end{equation}
($\mu,\nu=0,1,2,3;\, i=1,2,3;\, r=1,2$.)}
By analogy with \eqref{relweq0} and \eqref{relaveH0} the nonlinear parametric
quantization of gravity may be formulated with the following system
of equations:
\begin{equation}
\i \pd{{\psi}}{t} = \hat{\h} \psi
\label{relweq00}
\end{equation}
\begin{equation}
\ave{\hat{\HH}^\mu} = 0
\label{relaveH00}
\end{equation}
that {\em generates the nonlinear evolution of the {quantum state}
$\psi[\varphi^{{r}}; t)$ and {classical}
embedding variables $\vartheta^\mu(x^k,t)$}, using the operators
$\hat{\h}[\varphi^{{r}}, \hat{\varpi}_{{r}}, \vartheta^\nu, \dot{\vartheta}^\nu, N_\nu]$
and $\hat{\HH}^\mu[\varphi^{{r}}, \hat{\varpi}_{{r}}, \vartheta^\nu, \dot{\vartheta}^\nu, N_\nu; x)$
with $\hat{\varpi}_{{r}} := -\i\fd{}{\varphi^{{r}}}$.

It is an ongoing research program to further investigate the
quantization of gravity using \eqref{relweq00} and \eqref{relaveH00}
that effectively quantizes only two out of {the} six spatial metric components
as expected. In the meantime, it is instructive to investigate the implication
of the above quantization scheme via a
quantum cosmological approach. Therefore the
rest of this letter will be devoted to the parametric quantization
of a Friedmann universe with a massless scalar field based on
the much simpler set of equations  \eqref{relweq0} and \eqref{relaveH0}.
Such a cosmological model has the Lagrangian \cite{wang_2003, Blyth_Isham_1975}:
\begin{equation}\label{}
L(\phi, \dot\phi, R, \dot{R}, N) = -\frac{6 R }{N}\dot{R}^2 + \frac{R^3}{2N}\dot{\phi}^2  + 6 N K R
\end{equation}
in terms of the massless scaler
field $\phi(t)$, scale factor $R(t)$ and
lapse function $N(t)$ that enter into the
Robertson-Walker metric
\begin{equation}\label{RW}
g = -N^2\d t^2 + R^2 \sigma.
\end{equation}
Here, as usual,  $\sigma$ is the metric on the homogeneous and
isotropic 3-space of constant curvature $K$, with $K=1,0,-1$
corresponding to the closed, flat and open cases respectively.
Regarding  $R$ as the classical time variable we obtain from
\eqref{h} and \eqref{relh0} the following expressions:
\begin{equation}\label{}
\h =
6 R
\frac{\dot{R}^2}{N}
+
\frac{N {p}^2}{2R^3}  - 6 N K R
\end{equation}
\begin{equation}\label{}
\HH =
-6 R
\frac{\dot{R}^2}{N^2}
+
\frac{{p}^2}{2R^3}  - 6 K R
\end{equation}
where
\begin{align}\label{p}
  p &:= \pd{L}{\dot{\phi}} = \frac{R^3}{N}\dot{\phi}.
\end{align}
Comparing with the generic Lagrangian in \eqref{L} for $n=1$
we see that the current model has
the nonzero metric components and  potential as follows:
\begin{align}\label{}
\gamma_{00} &= -12 R,\quad \gamma_{11} = R^3 \\
V &=  - 6 K R.
\end{align}
By substituting $p \rightarrow \hat{p} := - \i \pd{}{\phi}$ into
$\h$ the operators
\begin{equation}\label{Kop}
\hat\h =
-\frac{N}{2R^3}\pd{^2}{\phi^2}
+
6 R
\frac{\dot{R}^2}{N}
- 6 N K R
\end{equation}
\begin{equation}\label{}
\hat\HH =
-\frac{1}{2R^3}\pd{^2}{\phi^2}
-
6 R
\frac{\dot{R}^2}{N^2}
- 6 K R
\end{equation}
are constructed that will act
on a wavefunction $\psi(\phi,t)$ (of weight $\frac12$
with respect to the 1-dimensional metric $\gamma_{11}$)
which is normalized according to:
\begin{equation}\label{}
{\ang{\psi}{\psi}} := \int_{-\infty}^\infty \!\!\abs{\psi(\phi,t)}^2 \d\phi = 1.
\end{equation}
On parametric quantization
the evolution equations for the wavefunction $\psi$ and classical variable $R(t)$
follow from \eqref{relweq0} and \eqref{relaveH0}
to be: \footnote{The term $6 R\frac{\dot{R}^2}{N^2}- 6 K R$ in $\hat\h$
is dropped as it contributes only to an overall phase in $\psi$. However this
term must be retained in applying \eqref{relaveH0}.}
\begin{equation}\label{sch}
\i \pd{{\psi}}{t}  = -\frac{N}{2R^3}\pd{^2{\psi}}{\phi^2}
\end{equation}
\begin{equation}\label{H0}
-6 R
\frac{\dot{R}^2}{N^2}
+
\frac{P^2}{2R^3}  - 6 K R = 0
\end{equation}
where
\begin{equation}\label{P}
P^2:=-\ave{\pd{^2}{\phi^2}}.
\end{equation}
The quantum evolution of the present cosmological model is therefore described by
the above system of {\em
nonlinear integro-partial differential
equations}.
Nonetheless, because of the absence of the ``potential'' term
(mass of the scalar being zero) it is possible to find analytic solutions by
regarding \eqref{sch} as the \sch{} equation of a free non-relativistic particle
with a time-varying mass. The analysis will be done in the $N=R$ gauge
in order to compare with
classical solutions \cite{Blyth_Isham_1975}. Thus \eqref{sch} is solved first to yield
\begin{equation}\label{gensol}
\psi(\phi,t) = \int_{-\infty}^{\infty} \frac{A(k)}{\sqrt{2\pi}}\, e^{\i [k \phi - F(t) k^2]} \d k
\end{equation}
where $A(k)$ is an arbitrary complex-valued function and
\begin{equation}\label{F}
F(t) := \int_{t_0}^{t} \frac{\d t'}{2R(t')^2}
\end{equation}
%for a chosen `initial time' $t_0$.
for a chosen reference time $t_0$. At this point we can see that
\begin{equation}\label{PP}
P^2 = \int_{-\infty}^{\infty} k^2 \abs{A(k)}^2 \d k
\end{equation}
which is a positive constant as anticipated with the free particle analogy.
Feeding this into \eqref{H0},
with suitable choice of the origin of the coordinate time $t$,
we immediately   obtain $R(t)$ to be
\begin{equation}\label{RR0}
{R(t)^2} = \left\{\begin{array}{cl}
\frac{{\abs{P}}\sin{2t}}{2\sqrt{3}} & (K=1)\\[5pt]
\frac{{\abs{P}} t}{\sqrt{3}} & (K=0)\\[5pt]
\frac{{\abs{P}} \sinh{2t}}{2\sqrt{3}} & (K=-1)
\end{array}\right. .
\end{equation}
Substituting these into \eqref{F} we find that
\begin{equation}\label{bc1}
F(t) = \left\{\begin{array}{cl}
%\frac{\sqrt{3}}{2P}\ln\left(\frac{\tan t}{\tan t_0}\right) & (K=1)\\[5pt]
%\frac{\sqrt{3}}{2P}\ln\left(\frac{t}{t_0}\right)& (K=0)\\[5pt]
%\frac{\sqrt{3}}{2P}\ln\left(\frac{\tanh t}{\tanh t_0}\right)& (K=-1)
\frac{\sqrt{3}}{2{\abs{P}}}\ln\frac{\tan t}{\tan t_0} & (K=1)\\[5pt]
\frac{\sqrt{3}}{2{\abs{P}}}\ln\frac{t}{t_0}& (K=0)\\[5pt]
\frac{\sqrt{3}}{2{\abs{P}}}\ln\frac{\tanh t}{\tanh t_0}& (K=-1)
\end{array}\right. .
\end{equation}

Solutions \eqref{RR0} are in fact same as in the classical case
if $P$ is replaced by the classical momentum $p$ defined in \eqref{p} which
is also a constant (of classical motion.) Unlike the classical case, though,
it is possible to envisage a quantum state with zero mean scalar momentum
($\ave{\hat{p}}=0$) but nonzero deviation of the scalar momentum
($\ave{\hat{p}^2}=P^2 > 0$). In this case, the evolution of the Friedmann
universe can be thought of as being ``purely quantum-driven.''
We conclude this letter by providing an explicit
example for such a scenario.
If at $t=t_0$ the wavefunction is given by the normalized gaussian packet
with a variance $\delta_0$
of the form:
\begin{equation}\label{}
\psi(\phi,t_0) = (2\pi)^{-\frac14}\delta_0{}^{-\frac12}\,
e^{- \frac{(\phi-\phi_0)^2}{4\delta_0^2}}\,e^{\i k_0 (\phi-\phi_0)}
\end{equation}
then
\begin{equation}\label{Ak}
A(k) = (2\pi)^{-\frac14}\sigma^{-\frac12}\,e^{- \frac{(k-k_0)^2}{4\sigma^2}}\,e^{-\i k \phi_0}
\end{equation}
where $k_0$ is the mean and $\sigma = \frac{1}{2\delta_0}$ is the variance of the
$k$-distribution.
Substituting \eqref{Ak} into \eqref{gensol} we obtain the following:
\begin{equation}\label{}
\psi(\phi,t) =
(2\pi)^{-\frac14}\left(\frac{\delta_0^2+\i F(t)}{\delta_0}\right)^{-\frac12}
e^{-\frac{\phi-\phi_0-2 k_0 F(t)}{4(\delta_0^4+F(t)^2)/\delta_0^2}}
\,
e^{\i\left[
k_0 (\phi-\phi_0)
+
\frac{\phi-\phi_0-2 k_0 F(t)}{4(\delta_0^4+F(t)^2)/F(t)}
-
k_0^2 F(t)
\right]}
\end{equation}
This is a moving wave packet with
the time-dependent standard deviation given by
\begin{equation}\label{delta}
\delta(t):=\abs{\frac{\delta_0^2+\i F(t)}{\delta_0}} = \sqrt{\frac{\delta_0^4+ F(t)^2}{\delta_0^2}}
{.}
\end{equation}
Clearly this yields the probabilistic density
\begin{equation}\label{}
\abs{\psi(\phi,t)}^2 =
\frac1{\sqrt{2\pi}\delta(t)}\,
e^{- \frac{(\phi-\phi_0-2 k_0 F(t))^2}{2\delta(t)^2}}.
\end{equation}
Furthermore, from \eqref{P} we have
\begin{equation}\label{PP1}
P^2 = k_0^2 + \sigma^2
{.}
\end{equation}
In particular, the choice of $x_0 = k_0 =0$ gives rise to the
``purely quantum-driven'' cosmological evolution discussed above.

It's my pleasure to thank Profs R W Tucker, C J Isham and M A H MacCallum
for fruitful conversations. Continued support from EPSRC and {BAE SYSTEMS}
enabling this work
is gratefully
acknowledged.

\end{document}